# Entropy Production in a Cell and Reversal of Entropy Flow as an Anticancer Therapy

**LiaoFu Luo**

**Laboratory of Theoretical Biophysics, Faculty of Science and Technology, Inner Mongolia University, Hohhot 010021, China;**

*Email address: lolfcm@mail.imu.edu.cn

**Abstract**

The entropy production rate of cancer cell is always higher than healthy cell under the case of no external field applied. Different entropy production between two kinds of cells determines the direction of entropy flow among cells. The entropy flow is the carrier of information flow. The entropy flow from cancer to healthy cell takes along the harmful information of cancerous cell, propagating its toxic action to healthy tissues. We demonstrate that a low-frequency and low-intensity electromagnetic field or ultrasound irradiation may increase the entropy production rate of a cell in normal tissue than that in cancer, consequently reverse the direction of entropy current between two kinds of cells. The modification of PH value of cells may also cause the reversal of the direction of entropy flow between healthy and cancerous cells. So, the biological tissue under the irradiation of electromagnetic field or ultrasound or under the appropriate change of cell acidity can avoid the propagation of harmful information from cancer cells. We suggest that this entropy mechanism possibly provides a basis for a novel approach to anticancer therapy.

Thermodynamic entropy is expressed by

$$S = k_B \ln W \tag{1}$$

where $W$ is the number of microscopic states which are related to a given macroscopic thermodynamic state and $k_B$ is the Boltzmann constant. Entropy is a measure of disorder. From general physical principles, Schrodinger first indicated that life should remain in a low-entropy state or "an organism feeds with negative entropy"[1]. This means that entropy production in an organism is canceled by the outward entropy flow so that the system remains in a highly ordered state of low entropy. However, following our point of view, negative entropy (or negentropy) is only the first half of the story. The living organism is a chemical engine in which a series of chemical reactions take place one by one in an appropriate sequence. Accordingly, the energy transfer in an organism in the normal state is so efficient that the entropy production is minimized. Minimal entropy production in a healthy cell is the second half of the story [2]. We shall compare qualitatively and demonstrate that the entropy production rate (or "entropy production" for short) of a healthy cell is lower than that of a cancerous cell if no external energy input [3-5]. However, when the appropriate external energy is applied to tissues, the rate of entropy production of normal cells may exceed that of cancerous cells. As an example, we shall discuss the entropy production of cells under irradiation of ultrasound and alternative electromagnetic fields. We shall prove the external energy applied to the body can reverse the direction of the entropy current. Since entropy current is the carrier of information current, as the direction of entropy current has been reversed the harmful effect brought about by the entropy flow from cancerous to healthy tissue will be blocked automatically. This gives a novel approach to anticancer therapy. The primary calculation and comparison of entropy production rates for healthy and cancerous cells were published originally in literatures [3-5]. In the article we shall summarize the recent advances in this study and give more complete discussions on the mechanism of the reversal of entropy-information flow between two kinds of cells which will be useful in the design of novel anti-cancer therapy.

**1. Entropy production in a living cell**

Entropy production is a thermodynamic quantity of fundamental importance for a living system since, following the second law of thermodynamics, entropy always increases for any non-equilibrium system. The entropy production $\sigma_s$ is the rate of entropy increase in unit volume. It can be proved that $\sigma_s$ contains five terms [6,7]:

1, $\sigma_s^{(1)}$    the thermal flux driven by a temperature difference;
2, $\sigma_s^{(2)}$    the diffusion current driven by a chemical potential gradient;
3, $\sigma_s^{(3)}$    the velocity gradient coupled with viscous stress;
4, $\sigma_s^{(4)}$    the chemical reaction rate driven by a Gibbs energy decrease (affinity);
5, $\sigma_s^{(5)}$    the dissipation due to the work completed by an external force field.

Formally, the rate is written as

$$\sigma_s = \sum_i \sigma_s^{(i)} \tag{2}$$

$$\sigma_s^{(1)} = \mathbf{j_q} \cdot \nabla \frac{1}{T} \tag{3}$$

$$\sigma_s^{(2)} = -\sum_\gamma \mathbf{j_\gamma} \cdot \nabla \frac{\mu_\gamma}{T} \quad (\mathbf{j_\gamma} = \rho_\gamma (\mathbf{v_\gamma} - \mathbf{V}), \quad \mathbf{V} = \sum_\gamma \frac{\rho_\gamma \mathbf{v_\gamma}}{\rho}) \tag{4}$$

$$\sigma_s^{(3)} = -\frac{1}{T} \sum_{ij} \partial_i V_j \Pi_{ji} \tag{5}$$

$$\sigma_s^{(4)} = \frac{1}{T} \sum_\delta J_\delta A_\delta \quad (A_\delta = -\sum_\gamma m_{\gamma\delta} \mu_\gamma) \tag{6}$$

$$\sigma_s^{(5)} = \frac{1}{T} \sum_\gamma \mathbf{j_\gamma} \cdot \mathbf{F_\gamma} \tag{7}$$

where $\mathbf{j_q}$ is the heat flux, $\mathbf{j_\gamma}$ - the diffusion flow of component γ, $\rho_\gamma$ - its concentration and $\mu_\gamma$ - its chemical potential, $\mathbf{v_\gamma}$ - its velocity and $\mathbf{V}$ the center of mass velocity of the cell fluid, $J_\delta$ is the number of the δ-th chemical reaction in unit volume and unit time, and $A_\delta$ - the affinity of the δ-th chemical reaction （or denoted as $\Delta G^0$ in literatures）, $\overset{\to\to}{\Pi}$ - the viscous stress tensor, describing the inner friction in the cellular fluid, and $F_\gamma$ - the external force acting on component $\gamma$ of unit mass.

**Calculation of entropy production in a cell without external field**

1. The entropy production rate in a cell due to heat flux driven by temperature difference is

$$\int \sigma_s^{(1)} d\tau = \int \mathbf{j}_q \cdot \nabla \frac{1}{T} d\tau$$

$$\approx \frac{1}{T^2} U_{cell} v_{therm} \frac{\delta T}{L_c} \tag{8}$$

where $\mathbf{j}_q$ is estimated by the product of internal energy density $U_{cell}$ and molecular thermal velocity $v_{therm}$, $\nabla \frac{1}{T} = -\frac{1}{T^2} \nabla T$ is related to temperature gradient, $L_c$ is a typical length in cell, and $\delta T$ - the temperature change in $L_c$. For normal cell the temperature is homogeneous, $\delta T=0$, so no entropy production due to thermal flux. However, for some cancers there is evidence on the increase of temperature[8]. One may take $\delta T=0.4^\circ C$ and $L_c$ = cell diameter in the estimate of the upper limit of entropy production for a cancerous cell.

2. The entropy production rate in a cell due to diffusion current driven by chemical potential gradient is

$$\int \sigma_s^{(2)} d\tau = -\int \sum_\gamma \mathbf{j}_\gamma \cdot \nabla \frac{\mu_\gamma}{T} d\tau$$

$$\approx \frac{1}{T} \frac{\overline{\mu} \overline{\rho} v_{therm}}{d_{memb}} \int_M d\tau$$

$$\approx \frac{1}{T} \frac{\overline{\mu} \overline{\rho} v_{therm}}{d_{memb}} V_{memb} \tag{9}$$

where $\mathbf{j}_\gamma$ is estimated by the product of concentration $\rho_\gamma$ (its mean denoted by $\overline{\rho}$) and thermal velocity $v_{therm}$ and $\overline{\mu}$ denotes the typical value of chemical potential $\mu_\gamma$, the gradient of which is approximated by $\overline{\mu}$ over a characteristic length $d_{memb}$. In fact, the gradient of chemical potential $\mu_\gamma$ in a cell occurs mainly in the region of packed membrane system (denoted as the integral over $M$ in (9)). Its volume $V_{memb}$ occupies a large portion of the cytoplasm.

3. The entropy production rate in a cell due to velocity gradient coupling with viscous stress is

$$\int \sigma_s^{(3)} d\tau = -\int \frac{1}{T} \nabla \mathbf{V} : \overrightarrow{\Pi} d\tau = -\frac{1}{T} \int \sum_{ij} \partial_i V_j \Pi_{ji} d\tau$$

$$= \frac{1}{T} \int \sum_{ij} \eta (\partial_i V_j)^2 d\tau \approx \frac{\bar{\eta}}{T} \int_{ecto} d\tau (\partial_i V_j)^2$$

$$\approx 4\pi \frac{\bar{\eta}}{T} \frac{V^2 r^2}{d_{ecto}} \qquad (10)$$

where the shear component of stress tenser $\mathbf{\Pi}_{ji}$ is approximately expressed by a linear law through viscosity coefficient $\eta$ (generalized Hooke's law): $-\eta \partial_i V_j$. In a living cell the cytosol occupies about 54% of the cell volume, surrounding many organelles[9] The internal cytoplasm which contains different granules is less viscous than the peripheral layer. The peripheral layer of the cytoplasm, the ectoplasm, behaves as a colloid. So the integral of eq (10) is approximated by the integral over cytoplasm layer (with thick $d_{ecto}$, $r$ means cell radius) and in this layer the average viscosity coefficient is denoted by $\bar{\eta}$.

4．The entropy production rate in a cell due to chemical reaction rate driven by affinity (Gibbs energy decreasing) is

$$\int \sigma_s^{(4)} d\tau = \int d\tau \frac{1}{T} \sum_\delta J_\delta A_\delta \qquad (11)$$

The structural change in cancerous cell causes the corresponding change in function[10,11]. Due to the decreased mitochondrial activity[12,13] the glycolysis, proteolytic and lipolytic processes have become the main energy sources of cancer cell where the extensive loss of skeletal muscle and adipose tissue occurs. The loss of adipose tissue is due to degradation of triglycerides, while the loss of skeletal muscles is due to an increased protein degradation[14]. The tumor products, like lipid mobilizing factor (LMF)[15] and proteolysis inducing factor (PIF)[16] have catabolic effects on host. The most important pathway of protein degradation, the ATP–ubiquitin, also leads to hyper-metabolism characteristic for parasitism[14]. All proteolytic and lipolytic processes, as catabolic reactions, contribute more entropy production in cancer than in normal cell. In the following we shall consider a simplified model. We shall study the main energy reaction –carbohydrate metabolism– only and suppose the ATP synthesis results from the oxidation of glucose, that is, from the full oxidation of glucose in normal cells and from the glycolysis in tumour tissue respectively.

In normal cells the full oxidation of 1 mole glucose will release 686 Kcal/mole and produces 31 mole (or 29.5 mole in the alternative pathway) ATP[17]. So the total Gibbs free energy decreasing is $\sum_\delta A_\delta$ (normal)=686-7.3×31×$f_n$ Kcal/mole in the respiratory chain of normal cell (the modifying factor $f_n$ stands for the dependence of standard free energy on pH value of cell environment in a normal cell). Instead, in cancerous cell the glycolysis is the main process where the total free energy release is 52 Kcal/mole and one mole glucose only produces 2 mole ATP. Correspondingly, the Gibbs free energy decreasing is $\sum_\delta A_\delta$ (cancer)=52-7.3×2×$f_c$ Kcal/mole in the cancer cell (the modifying factor $f_c$ comes from the dependence of standard free energy on pH value of cancerous cell). The typical pH value for normal cell is 7.4, while for cancer it is 6.9. We assume the factor $f_n$-1 or $f_c$-1 proportional to the difference between pH value and 7, namely, $f_n$=1.68 and $f_c$=0.83. So,

$$\sum_\delta A_\delta \text{ (normal)}=305.8 \text{ Kcal/mole}, \quad \sum_\delta A_\delta \text{ (cancer)}=39.9 \text{ Kcal/mole}. \tag{12}$$

On the other hand, the glucose consumption in a cancerous cell is much higher than in a healthy cell. It was demonstrated through positron emission tomography scanning that tumor cells absorb more glucose than normal cells. Cancer cells metabolise glucose at a rate of approximately 20 times that of normal tissue. From the comparison of ATP molecule number produced in glycolysis and that in glucose oxidation of respiratory chain we estimate the average rate of the oxidation of glucose in unit volume of a cancerous cell ($J_c$) is 15.5 times higher than that in a normal cell ($J_n$).

Lehninger reported that the biological synthesis in one *E coli* cell needs energy of 2.4 million ATP molecules per second[18]. Assuming an adult consumes 40Kg ATP in 24 hours we estimate the ATP consumption is $1.2 \times 10^6 \sim 1 \times 10^8$ per second for different sizes of cells in human body. This means

$$\int J_n d\tau = 3.9 \times 10^4 /s \quad \text{to} \quad 3.2 \times 10^6 /s$$

$$\int J_c d\tau = 15.5 \int J_n d\tau \tag{13}$$

By use of Eq (12)(13) we estimate the rates of entropy production in a cell are

$$\int \sigma_s^{(4)} d\tau \text{ (normal)} = 2.7 \times 10^{-9} \text{ erg/sec} \cdot \text{deg} \quad \text{(for cell with volume 200μ}^3\text{)}$$

$$\text{to} = 2.2 \times 10^{-7} \text{ erg/sec} \cdot \text{deg (for cell with volume 15000μ}^3\text{)} \tag{14a}$$

$$\int \sigma_s^{(4)} d\tau \text{ (cancer)} = 2 \int \sigma_s^{(4)} d\tau \text{ (normal)} \tag{14b}$$

(If the pH-dependence of standard free energy is not considered, $f_n = f_c =1$, then the factor 2 in the right hand side of eq (14b) should be replaced by 1.3 ). So, using the above-mentioned simplified model we estimate the rate of entropy production in a cancerous cell is about two times higher than that in a healthy cell from the main energy reaction.

**Numerical estimates of entropy production for a cell without external energy input**

In tissues of the human body, with the exception of some nerve cells, the volume of cells varies between 200μ$^3$ and 15000μ$^3$ (diameter 7 to 30 μ)[20]. So in the numerical estimates the typical values 2r=30μm, $M_{cell}$=1.5$\times 10^{-8}$ gm, and $\bar{\rho}$=1gm/cm$^3$ are taken. Referred to the mean velocity of small molecules in cytoplasm we take $v_{therm}$= 5$\times 10^{-3}$ cm/s. The internal energy $U_{cell}$ is estimated smaller than $3 \times 10^{-22}$ *erg* by comparing a cell with the water of same mass. In eq (9) the typical chemical potential $\bar{\mu}$ is supposed to be μ$_{ATP}$=7.3Kcal/mole=0.12$\times 10^{-6}$ erg/gram;

simultaneously the volume of packed membrane system $V_{memb}$= 7000$\mu^3$ and characteristic length $d_{memb}$ =0.01$\mu$ are assumed. In calculating eq(10) the velocity $V$ is estimated $3 \times 10^{-4} cm/s$ through comparison with the mean velocity of 70 kD protein in *E coli* cytoplasm[19], the thickness of peripheral layer of the cytoplasm $d_{ecto}$ is supposed to be 0.2 r. The average viscosity coefficient $\bar{\eta}$ ==10 $\eta$(water)=0.114 (in CGS unit system) is taken which is comparable with $\eta$(*Amoeba dubia*)= 2.24$\eta$(water) and $\eta$(*Slime molds*)= (10 - 20) $\eta$(water) at room temperature[20]. The final results of numerical estimate are

$$\int \sigma_s^{(1)} d\tau \approx 2.2 \times 10^{-27} \text{ erg/degree sec} \quad \text{(for some cancerous cell)}$$

$$\approx 0 \text{ (for healthy cell)}$$

$$\int \sigma_s^{(2)} d\tau \approx 1.1 \times 10^{-22} \text{ erg/degree sec}$$

$$\int \sigma_s^{(3)} d\tau \approx 3.1 \times 10^{-12} \text{ erg/degree sec} \quad (15)$$

$\int \sigma_s^{(4)} d\tau$ has been given by eq(14). We find the rate due to chemical reaction ($i = 4$) much higher than other three terms ($i = 1,2,3$). Simultaneously the chemical reaction contributes to the entropy production differently between healthy and cancerous cells. Thus, we obtain an important conclusion: the rate of total entropy production in a cancerous cell is generally higher than that in a normal cell in case of no external energy input. Although there is some arbitrariness in above parameter choice, the conclusion holds independently of the choice. Our result is consistent with the point of the minimal entropy production theorem.

Non-equilibrium statistical physics affords an important clue for the understanding of the self-organization phenomena of living bodies. Prigogine proved that, in the linear range of an irreversible process in non-equilibrium thermodynamics, the entropy production rate of a steady state always takes up a minimum[7]. In his proof the condition of local equilibrium and the stability of the local equilibrium should be assumed. However, this condition is not satisfied for a general nonequilibrium system. On the other hand, Onsager's reciprocity relation between flow and force has also been used in the proof, which depends on the invariance under time reflection (the microscopic reversibility). But for living body as a chiral system the invariance under time reversal is broken. So, whether the minimum entropy production holds in a chiral system such as living body should be re-examined. Recently, we indicated that, if the local equilibrium and its stability are valid for each step of the process then the minimum entropy production can be proved not only for linear region, but also for some nonlinear regions where the force is small but the corresponding flow not small. In this proof the Onsager relation is not required[2]. The living organism is a chemical engine in which a series of chemical reactions take place one by one in an appropriate sequence. Accordingly, the energy transfer in an organism in the normal state is so efficient that the entropy production is minimized. The physical picture is as follows. The entropy production is a function of a group of parameters which can be described by a hyper-surface in the parameter's space. It is a rugged landscape essentially. For a living body in normal state, under the action of 'force' $\frac{d}{dt}\int \sigma_s d\tau \leq 0$, the system will tend to one of the valleys of the landscape. It means that the condition of stable local equilibrium seems satisfied and the minimum entropy production holds for normal cells. The point is consistent with the comparative analysis of entropy production in healthy and cancerous cells given above.

**2. Relation between entropy and information quantity**

To clarify the Shannon information quantity, let us consider the information conveyed by the symbols $s_i$ of a source $S$ $\{s_i\}$, the probability of $s_i$ being $p_i$. The information quantity represents how much information is gained by knowing that $S$ has definitely emitted the $i$-th symbol $s_i$; this also represents our prior uncertainty as to whether $s_i$ will be emitted, and our surprise on learning that it has been emitted. Thus, the concept of information quantity is essentially similar to the description of entropy, and this explains why we usually refer to the information quantity as information entropy. Mathematically, for a system with a given distribution of probable states, the Shannon information quantity is defined by

$$I = -\sum_i p_i \log_2 p_i \tag{16}$$

where $p_i$ is the probability of occurrence of the $i$-th state. For an equiprobable distribution of $N$ states, $p_i = \frac{1}{N}$ and one has

$$I = \log_2 N = \frac{1}{\ln 2}\ln N \tag{17}$$

Comparing (17) with (1) one obtains the information quantity $I$ proportional to the thermodynamic entropy $S$. Note that the proportionality exists between the information quantity and entropy even for a non-equiprobable distribution of states. In fact, when Shannon obtained equation (16) he named it as information entropy under the suggestion of Szilard that the expression is close to Boltzmann's entropy. But, is there any difference between information quantity and thermodynamic entropy apart from some constant factor? Our point is the essential difference between two quantities consists in the number of states in their definition. Since the number of microscopic states $W$ in eq (1) is very large while the number of states $N$ in Shannon information is generally much smaller, the information quantity should be regarded as the projection of thermodynamic entropy from the microscopic phase-space to the subspace spanned by $N$ macroscopic states.

As we know, the thermodynamic entropy of a cancerous cell is different from that of a normal cell due to the more disordered structure of the cancerous cell. Correspondingly, due to the different sub-cell structures and physiological states between two kinds of cells the information inherent in a cancerous cell is different from that in a normal cell. The information quantities in cancerous and normal cells are both described by equation (16), but they have different distributions of $\{p_i\}$, $p_i$(cancer) $\neq p_i$ (normal) ($i = 1,…,N$). ($p_i$ is the probability of the $i$-th chemical, morphological, structural or physiological state of the cell). The particular set of $\{p_i\}$ in a cancerous cell, $\{p_i$(cancer)$\}$, deviates from the normal value and reflects the particular bias of the states in a tumor. While the information in a healthy cell is defined by the set of $\{p_i\}= \{p_i$(normal), the corresponding information quantity in a cancerous cell based on $\{p_i$(cancer)$\}$ is called harmful information.

The difference between information quantity and thermodynamic entropy can be formulated by mathematical equations.

Set the probability density in phase space of microscopic states denoted by **ρ** and that in macroscopic state space by **P**. The thermodynamic entropy is defined by **ρ** while the **i**nformation quantity is defined by **P**. The relation between **ρ** and **P** is

$$\mathbf{P} = \text{Proj } \boldsymbol{\rho} \tag{18}$$

Here Proj means projection operator. As is well known, **ρ** obeys Liouville equation[21]，

$$\frac{\partial \rho(q_1, p_1, \ldots, q_s, p_s, t)}{\partial t} = -iL\rho(q_1, p_1, \ldots, q_s, p_s, t) \tag{19}$$

($q_i$ and $p_i$ are canonical coordinate and momentum for the *i*-th atom). L is Liouville operator expressed in the form of Poissin bracket. However, **P** obeys Kolmogorov equation. Suppose the probability of macroscopic states is defined by a group of variables of state *a* and configuration **x**. Under some general conditions on transition function $P(a', \mathbf{x}', t'; a, \mathbf{x}, t)$ Kolmogorov proved that $P(a, \mathbf{x}, t) = P(a, \mathbf{x}, t; a_0, \mathbf{x}_0, t_0)$ obeys [22]

$$\frac{\partial P(a, \mathbf{x}, t)}{\partial t} = -\frac{\partial}{\partial a}(K_1 P(a, \mathbf{x}, t)) + \frac{1}{2}\frac{\partial^2}{\partial a^2}(K_2 P(a, \mathbf{x}, t)) \\ -\nabla \bullet (\mathbf{Q}_1 P(a, \mathbf{x}, t)) + \frac{1}{2}\nabla^2(Q_2 P(a, \mathbf{x}, t)) \tag{20}$$

where $K_1$ and $\mathbf{Q}_1$ are the first-order jump moment with respect to variable *a* and **x** and $K_2$ and $\mathbf{Q}_2$ are the second-order jump moment with respect to variable *a* and **x** respectively. Eq (20) is much different from Liouville equation(19). So two probability distributions obey quite different equations. For thermodynamic entropy one should proceed from Liouville equation, while for Shannon information the starting point is Kolmogorov equation. The formulas for information density function derived from Kolmogorov equation was firstly given by Xing [23].

The thermodynamic entropy of a system (a normal cell, a cancerous cell, etc.) changes with time, obeying the continuity equation (entropy balance equation) [6]:

$$\frac{dS}{dt} = \int \sigma_s d\tau + \text{(net rate of entropy flow through boundary)} \\ = \int \sigma_s d\tau + \text{(entropy flow rate in)} - \text{(entropy flow rate out)} \tag{21}$$

or its differential form

$$\frac{\partial(\rho s)}{\partial t} + \nabla \cdot \mathbf{j_s} = \sigma_s$$

where *s* is entropy of unit mass and $\mathbf{j_s}$ the entropy flow. Following the second law of thermodynamics, the entropy production is always positive. Only when the entropy production is canceled by the outward entropy flow can the system remain in an ordered low-entropy state. The entropy flow is expressed by[6]

$$\mathbf{j_s} = \rho s \mathbf{V} + \frac{1}{T}\mathbf{j_q} - \frac{1}{T}\sum_{\gamma} \mu_\gamma \mathbf{j}_\gamma \qquad (22)$$

It contains three terms: the convection term of entropy, the conduction term related to transport of heat, and the conduction term related to transport of matter.

From the comparison of the definitions of thermodynamic entropy and information quantity given above, it is easy to understand the relation between information flow and entropy flow. Since the information quantity is the projection of the thermodynamic entropy, the entropy flow should be the carrier of the information flow. So, accompanying the entropy flow there should be certain information transport. Thus, the entropy flow from a normal to a cancerous cell carries the information of the healthy cell, while the entropy flow in the opposite direction carries the harmful information of the cancerous cell.

From eq.(22) we can easily determine the direction of entropy flow. The first term in eq.(22) involves the entropy transport from a site of high entropy density to one of low entropy density that accompanies convection movement. Due to the higher entropy production in cancer, the convection of entropy proceeds in the direction from cancerous to healthy cells. The second term is in the direction of temperature gradient. Due to the higher temperature that a cancer cell may have, the conduction of entropy related to heat transport also proceeds in the direction from cancerous to healthy cells though this is generally a small term. The third term of entropy flow, the conduction of entropy related to the transport of matter, is in the opposite direction to the matter transport. If the flow of matter transport is mainly from healthy to cancerous cells (this is the case for many cancers), then the conduction term is again in the direction from cancerous to healthy cells. Therefore, we conclude that the entropy always flows from cancerous cells to healthy ones if no special therapeutic design has been introduced. Since the entropy flow is the carrier of information flow this would lead to the propagation of harmful information from cancerous cells.

**Changing entropy flow direction through adjusting cell pH**

How to change the direction of entropy flow? An important approach is to change the pH value of cells. As stated above the entropy production rate due to chemical reaction is dependent on the cell's pH. Consider the oxidation of glucose to synthesize ATP molecule and study how the free energy change dependent on the pH value of cell. For a reaction

$$aA + bB \rightleftharpoons cC + dD$$

the equilibrium constant is

$$K_{eq} = \frac{[C]^c_{eq}[D]^d_{eq}}{[A]^a_{eq}[B]^b_{eq}} \qquad (23)$$

and the standard free energy is related to equilibrium constant through

$$\Delta G^0 = -2.3RT \log K_{eq} \qquad (24)$$

where $R$ is gas constant[17]. The standard free energy change $\Delta G^0$ at pH 7 is denoted by $\Delta G^{0*}$. Le Chatelier's principle states that if a stress is applied to a reaction at equilibrium the equilibrium will be displaced in the direction that relieves the stress. Under basicity environment (pH>7) the equilibrium of glucose oxidation reaction will be displaced in the direction that increases acidity and therefore the equilibrium constant increases. It leads to $\Delta G^0 < \Delta G^{0*}$. Oppositely, under acidity environment (pH<7) the equilibrium of glucose oxidation reaction will be displaced in the direction that decreases acidity and therefore the equilibrium constant decreases. It leads to $\Delta G^0 > \Delta G^{0*}$. For the reaction of ATP synthesis from glucose oxidation the standard free energy change is $\Delta G^{0*}$ =686-7.3×31 Kcal/mole in the respiratory chain or 52-7.3×2 Kcal/mole in the glycolysis. More basicity leads to standard free energy $\Delta G^0$ decreasing and more acidity leads to standard free energy $\Delta G^0$ increasing. Therefore the entropy production $\int \sigma_s^{(4)} d\tau$ (or the Gibbs free energy decreasing $\sum_\delta A_\delta$ ) is lowered when the cell pH grows, and raised when the cell pH drops. This means that the decrease of the acidity for cancerous cells and/or the increase of the acidity for normal cells can change the relative entropy production rate of two kinds of cells and cause the reversal of entropy flow.

Another approach to the change of entropy flow is to reduce the transport of matter from normal to cancerous cells. Angiogenesis as a therapeutic target has recently been widely discussed. It has become apparent that the targeted destruction of the established tumour vasculature is an avenue leading to exciting therapeutic opportunities[24]. From our point of view, the modulation of angiogenesis and the lowering of the glucose supply to the cancerous cells are favourable for decreasing entropy production and reversing the direction of entropy flow. On the other hand, increasing of the housing temperature can reverse the direction of entropy flow. It lowers the temperature gradient in tumorous cells and reduces the rate of their entropy production that comes from the heat flux. Both factors are of benefit in cancer therapy. However, the mathematical estimate has indicated that the entropy production due to heat is only a very small fraction of the total entropy production. So the effectiveness of hyperthermia used as therapy is very low.

So far we have not considered the fifth term of entropy production of equation (2). In case of external energy input, the entropy production rate of a normal cell may exceed the cancer and therefore the direction of entropy flow can be reversed. This physical principle leads to an interesting conclusion. Under appropriate external energy input the entropy flow direction will be reversed; it directs from normal to cancerous cells, carrying the information of healthy tissues, blocking the propagation of harmful information of cancerous cells. This provides opportunities for anticancer therapy. Now we shall consider two cases of external energy input.

**3. Electro-magnetic field causes reversal of entropy flow**

Suppose alternating electric field of angular frequency $\omega$ is applied on the cell. The cell is modelled as a spheroidal conductive dielectric medium with two structural parts – cell membrane and cytoplasm. Every cell structural part can be described by two parameters – permittivity $\varepsilon_m$ or $\varepsilon_p$ and conductivity $\sigma_m$ or $\sigma_p$ for membrane or for cytoplasm respectively. The cell membrane has specific capacitance which is proportional to the ratio of the membrane permittivity to the membrane thickness. Thus, the membrane can be modelled as a capacity (with capacitance $C_m$) and a resistance ($R_m$) in series and the cytoplasm simply as a resistance ($R_p$) in the equivalent circuit. We shall neglect the dispersion related to the molecular properties of cytoplasm. Simultaneously, we shall neglect the magnetic power dissipation since it is significant only in materials with high magnetic permeability. Based on the above model we are able to deduce the electromagnetic entropy production of a cell through calculating the rate of Joule loss ($Q_m$ for membrane and $Q_p$ for cytoplasm) in the electric circuit.

$$Q_m = \frac{(E_m d)^2 R_m}{R_m^2 + \dfrac{1}{C_m^2 \omega^2}} \tag{25}$$

$$Q_p = \frac{(E_p D)^2}{R_p} \tag{26}$$

where $E_m$ and $E_p$ are macroscopic electric field (effective value, $\dfrac{1}{\sqrt{2}} \times$ amplitude) in membrane and cytoplasm respectively, $d$ is the thickness of membrane in the direction of electric field and $D$ the dimension of cytoplasm in that direction. The macroscopic electric field $\mathbf{E_m}$ and $\mathbf{E_p}$ obeys continuity condition

$$\begin{aligned} \mathbf{n} \cdot (\varepsilon'_m \mathbf{E}_m - \varepsilon'_p \mathbf{E}_p) &= 0 \\ \mathbf{n} \times (\mathbf{E}_m - \mathbf{E}_p) &= 0 \end{aligned} \tag{27}$$

where $\varepsilon'$ is complex permittivity and $\mathbf{n}$ is a unit vector normal to boundary. It leads to

$$\frac{E_p}{E_m} = \sqrt{\left(\frac{\varepsilon_m}{\varepsilon_p}\right)^2 \cos^2 \theta + \sin^2 \theta} \tag{28}$$

for given angle $\theta$ between the incident electric field and the normal. The macroscopic field in dielectric medium is related to the applied electric field $E_a$ through[25]

$$E_p = \frac{3}{\varepsilon_p + 2} E_a \tag{29}$$

Inserting (28)(29) into (25) (26) we obtain

$$Q_p = \frac{9 E_a^2 D^2}{(\varepsilon_p + 2)^2 R_p} \tag{30}$$

$$Q_m = \frac{9 E_a^2 d^2 R_m}{(\varepsilon_p + 2)^2 \left(R_m^2 + \dfrac{1}{C_m^2 \omega^2}\right)\left(\left(\dfrac{\varepsilon_m}{\varepsilon_p}\right)^2 \cos^2 \theta + \sin^2 \theta\right)} \tag{31}$$

The membrane loss $Q_m$ is dependent on field frequency, which behaves $\omega^2$ at low frequency and approaches

$$Q_m(high) = \frac{9E_a^2 d^2}{(\varepsilon_p + 2)^2 (\frac{\varepsilon_m^2}{\varepsilon_p^2}\cos^2\theta + \sin^2\theta)R_m} \tag{31.1}$$

at high frequency limit. For the incident field of large θ the Joule loss is relatively low due to its short path in cell. Only the small θ component makes the important contribution to the loss. Comparing $Q_m$ with $Q_p$ at small θ we estimate

$$\frac{Q_m}{Q_p} \leq \frac{\varepsilon_p^2 d^2 R_p}{\varepsilon_m^2 D^2 R_m} = \frac{\varepsilon_p^2 \sigma_m d}{\varepsilon_m^2 \sigma_p D} \tag{32}$$

for alternating electric field of different frequencies. Taking[26]

$$\varepsilon_p \sim 60, \quad \varepsilon_m \sim 10, \quad \sigma_p \sim 1\text{S/m}, \quad \sigma_m \sim 10^{-5}\text{S/m}, \quad \frac{d}{D} \sim 10^{-3} \tag{33}$$

we find membrane loss several order lower than cytoplasm loss. In the following calculation we shall neglect it.

From Eq (7) the entropy production $\int \sigma_s^{(5)} d\tau$ is related to the coupling of external force and diffusion current. The work completed by external force field includes conservative part which is equal to the coupling of force and center-of-mass velocity and contributes to the increase of mechanical energy of the system, and dissipative part which is equal to the coupling of force and diffusion velocity and contributes to the entropy or thermal energy of the system. So, the entropy production rate $\int \sigma_s^{(5)} d\tau$ due to alternating electric field can be calculated through Joule loss $Q_p$ and $Q_m$. As neglecting $Q_m$ we have

$$\int \sigma_s^{(5)} d\tau = \frac{Q_p}{T} = \frac{9E_a^2 D^2}{(\varepsilon_p + 2)^2 R_p T} \tag{34}$$

where $D$ can be approximated by cell diameter and $R_p \approx \frac{1}{\sigma_p D}$. Taking

$$T = 310K, \quad \varepsilon_p \sim 60, \quad D = 30\mu m, \sigma_p \sim 1\text{S/m}, \quad E_a = 3\text{volt/cm} \tag{35}$$

we estimate entropy production $\int \sigma_s^{(5)} d\tau \approx 0.2 \times 10^{-6}$ erg/sec·degree, which is in the same order of cell entropy production without external field (eq 14).

Now we consider the difference of entropy production between cancerous and healthy cells. Define $(\varepsilon_p + 2)^{-2} \sigma_p$ of a cell as its entropy production threshold (EPT1).

$$\text{EPT1(cancer)} = (\varepsilon_p(cancer) + 2)^{-2} \sigma_p(cancer)$$

$$\text{EPT1(healthy)} = (\varepsilon_p(healthy) + 2)^{-2} \sigma_p(healthy) \tag{36}$$

From (34) one has

$$\frac{\int \sigma_s^{(5)} d\tau (cancer)}{\int \sigma_s^{(5)} d\tau (healthy)} = \frac{EPT1(cancer)}{EPT1(healthy)}$$
$$= \frac{(\varepsilon_p(healthy)+2)^2}{(\varepsilon_p(cancer)+2)^2} \frac{\sigma_p(cancer)}{\sigma_p(healthy)} \quad (37)$$

When the EPT1 of cancerous cell is lower than that of healthy cell then the entropy flow (information flow) direction will be reversed as the organism irradiated under low intensity alternating electric field (with field strength several volts/cm). So EPT1 is a threshold value of a cell which describes whether the electric field irradiation on it may have physiotherapeutic effect of anticancer.

The dielectric properties of normal and malignant tissue have been measured by several authors. Barsamian, Reid and Thornton [27] studied dielectric permittivity ($\varepsilon_p$) and specific electro-conductivity ($\sigma_p$) of cells of D tryoni at different stages of development for normal cells and for cells after carcinogen 20-MC treated. For pupae, they obtained normal $\sigma_p=3.05 \times 10^{-5}$ (ohm mm)$^{-1}$, cancerous $\sigma_p=5.21 \times 10^{-5}$ (ohm mm)$^{-1}$, and normal $\varepsilon_p=0.56 \times 10^5$, cancerous $\varepsilon_p=0.88 \times 10^5$. The EPT1 of cancerous cell is lower than that of healthy cell by a factor 0.69. For adult, they obtained normal $\sigma_p=3.13 \times 10^{-5}$ (ohm mm)$^{-1}$, cancerous $\sigma_p=5.49 \times 10^{-5}$ (ohm mm)$^{-1}$, and normal $\varepsilon_p=0.33 \times 10^5$, cancerous $\varepsilon_p=0.59 \times 10^5$. The ratio of EPT1 for cancer to normal is 0.55. Polevaka et al [26] obtained dielectric parameters of human white blood cells $\sigma_p=1.31$ S/m for normal B cells and $\sigma_p=0.48$ S/m for malignant B cell line Farage by fixing parameter $\varepsilon_p=60$. The ratio of EPT1 for cancer to normal is 0.37. Sha et al，[28] summarized the dielectric measurements of breast normal and malignant tissue and indicated data inconsistency at low frequency. But at high frequency they argued that malignant tissues have higher mean permittivity and conductivity values than those of normal. Haemmerich et al [29] indicated the conductivity of normal liver tissue $\sigma_p=1.26$ mS/cm at 10 Hz and $\sigma_p=4.61$ mS/cm at 1MHz, while for tumour $\sigma_p=2.69$ mS/cm at 10 Hz and $\sigma_p=5.23$ mS/cm at 1MHz. But no permittivity data was provided.

Now we consider alternating magnetic field $\mathbf{B} = \mathbf{B}_0 \sin \omega t$ applied on the cell[44]. The cell is again modelled as a spheroidal medium with two structural parts – cell membrane and cytoplasm. Most of the organisms are diamagnetic material. Only a few of the tissues (Fe-containing haemoglobin) are paramagnetic material. The magnetic susceptibility χ is generally a small quantity (about -$10^{-5}$). So one may neglect the difference between magnetic permittivity μ in cytoplasm and that in membrane. The magnetic field $\mathbf{B}$ satisfies continuity condition $\mathbf{B}_m = \mathbf{B}_p$ in the boundary of two phases as $\mu_m = \mu_p$ since

$$\mathbf{n} \cdot (\mathbf{B}_m - \mathbf{B}_p) = 0$$

$$\mathbf{n} \times (\frac{\mathbf{B}_m}{\mu_m} - \frac{\mathbf{B}_p}{\mu_p}) = 0 \tag{38}$$

Following Faraday's law one has electromotive force **E** satisfying $\oint \mathbf{E} \cdot d\mathbf{l} = -\frac{\omega}{c} AB_0 \cos \omega t$ in vacuum where $A$ is the cross section area of the circuit. Taking the difference between macroscopic electric field and applied electric field in dielectric medium into account we have the dissipative electric power generated by induction current (averaged over an alternating current period)

$$Q_p = \frac{\omega^2 A^2 B_0^2}{2c^2 R_p} (\frac{3}{\varepsilon_p + 2})^2 \tag{39}$$

for cytoplasm in the deduction of which eq (29) has been used. The factor $(\frac{3}{\varepsilon_p + 2})^2$ describes the polarization effect of cytoplasm. If the similar relation between the applied electric field and the macroscopic electric field is assumed in membrane then one has the dissipative electric power

$$Q_m = \frac{\omega^2 A^2 B_0^2 R_m}{2c^2 (R_m^2 + \frac{1}{C_m^2 \omega^2})} (\frac{3}{\varepsilon_m + 2})^2 \tag{40}$$

for membrane. Different from electric field, the dissipative power is dependent of field frequency. In high frequency limit, both the membrane loss $Q_m$ and cytoplasm loss $Q_p$ depend on field frequency as $\omega^2$ and one has

$$\frac{Q_m}{Q_p} = \frac{R_p (\varepsilon_p + 2)^2}{R_m (\varepsilon_m + 2)^2} \simeq \frac{\sigma_m (\varepsilon_p + 2)^2 D}{\sigma_p (\varepsilon_m + 2)^2 d} \quad (\omega >> \frac{1}{R_m C_m}) \tag{41}$$

($D$ - cell diameter, $d$ – membrane thickness). At low frequency,

$$\frac{Q_m}{Q_p} \approx \omega^2 C_m^2 R_m R_p \frac{(\varepsilon_p + 2)^2}{(\varepsilon_m + 2)^2} \quad (\omega << \frac{1}{R_m C_m}) \tag{41.1}$$

The oscillatory time $R_m C_m = \frac{\varepsilon_m}{4\pi \sigma_m}$ is estimated to be $10^{-5}$ s if $\varepsilon_m = 10$ and $\sigma_m = 10^{-5}$ S/m are taken. The entropy production rate in a cell due to alternative magnetic field

$$\int \sigma_s^{(5)} d\tau = \frac{Q_p + Q_m}{T} = \frac{9\omega^2 A^2 B_0^2 k}{2c^2 (\varepsilon_p + 2)^2 R_p T}$$

$$\cong 2(\frac{3\pi}{8})^3 k \frac{\omega^2 B_0^2 D^5 \sigma_p}{c^2 (\varepsilon_p + 2)^2 T} \quad (k = 1 + \frac{Q_m}{Q_p}) \tag{42}$$

The factor k takes a value larger than but near 1, for example, k=1.26 for the case of ω >>($R_m C_m$)$^{-1}$ and parameters given by eq (33). Consider the magnetic field of frequency 1MHz. In this frequency range the ratio of $Q_m$ to $Q_p$ is given by eq (41). By using (33) and

$$T = 310K, \quad \varepsilon_p \sim 60, \quad D = 30\mu m, \sigma_p \sim 1S/m,$$

$$\omega = 2\pi\nu = 2\pi \times 10^6 s^{-1}, \quad B_0 = 10^4 Gs \tag{43}$$

we estimate $\int \sigma_s^{(5)} d\tau = 3.2 \times 10^{-8} erg/\deg \cdot \sec$ which is in the same order of the cell entropy production without external field (eq 14).

The difference of entropy production between cancerous and healthy cells can be deduced from eqs (39) and (40). Define entropy production threshold (EPT2)

$$EPT2 = \frac{\sigma_p}{(\varepsilon_p + 2)^2} + \frac{D}{d}\frac{\sigma_m}{(\varepsilon_m + 2)^2} \tag{44}$$

We have

$$\frac{\int \sigma_s^{(5)} d\tau(cancer)}{\int \sigma_s^{(5)} d\tau(healthy)} = \frac{EPT2(cancer)}{EPT2(healthy)} \tag{45}$$

for magnetic field of moderate frequency higher than 0.1 MHz, again the same formula as eq(37) for electric field. Note that the second term of EPT2 is the modification to EPT1 (the first term of eq (44)) and the ratio of the former to latter is 0.26 as the parameter choice eq (33) is assumed.

So, under the alternating electric field the additional entropy production for normal cell is higher than that for cancer as EPT1(cancer) < EPT1(normal); likewise, under the alternating magnetic field the additional entropy production for normal cell is higher than that for cancer as EPT2(cancer) < EPT2(normal). In previous discussions we estimated that for a cell in case of no force field applied, the entropy production is $10^{-7}$ erg/degree/sec or lower. So, 3 volt/cm low frequency alternating electric field or $10^4 Gs$ alternating magnetic field with moderate frequency 1MHz can induce the additional entropy production the order of which is comparable with the entropy production for a cell without applied field. The additional entropy production will reverse the direction of entropy flow and avoid the propagation of harmful information of cancer to surrounding normal tissue. We suggest that this entropy mechanism possibly provides a basis for a novel approach to the physiotherapy of anticancer. Note that the new approach is non-damaging, less-invasive for human body in clinical application.

Note 1. We have used continuous wave electromagnetic field to investigate the entropy production. The discussion can easily be generalized to other waveforms. Recently the nanosecond pulsed electric field effects on human cells have attracted great attention of investigators by its potential applications. These pulses can generate extremely high power, but because they are so short, they have a low energy density and produce negligible heating effects. The observed effects of nanosecond pulsed electric field on biological cells (for example, the apoptosis induction[37]) can be dealt with partly in the framework of the present work.

2. The interaction between radiofrequency fields and biological system is generally related to thermal mechanism (the magnitude of temperature increase from the exposure). For example, the magnetic fluid hyperthermia was proposed as an anticancer therapy system[30]. The thermal mechanism is measured by the specific absorption rate (SAR) with threshold value about 1W/kg (or $1.5 \times 10^{-4}$ erg/s for a typical cell) which corresponds roughly to the basal metabolic rate of humans [31]. The electromagnetic field entropy production studied in this article has typical value of $10^{-7}$ erg/degree/sec, lower than the threshold, so the mechanism proposed in this article is mainly nonthermal.

3. The relation between macroscopic field $\mathbf{E}_p$ in dielectric medium and applied electric field $\mathbf{E}_a$, Eq (29), is deduced as follows:

$$\mathbf{E}_p = \mathbf{E}_a + \mathbf{E}_{self}$$

where the self-field of homogeneously polarized sphere is given by

$$\mathbf{E}_{self} = -\frac{4\pi}{3}\mathbf{P} \tag{46}$$

So,

$$\mathbf{E}_a = \mathbf{E}_p + \frac{4\pi}{3}\mathbf{P} = \mathbf{E}_p + \frac{\varepsilon - 1}{3}\mathbf{E}_p = \frac{\varepsilon + 2}{3}\mathbf{E}_p$$

The deduction is rigorous only when $\omega\tau \ll 1$. Here $\tau$ means the relaxation time of medium polarization. Consider a dipolar molecular of radius $a$ moving in a continuous viscous fluid with viscosity $\eta$, Debye deduced[25]

$$\tau = \frac{4\pi\eta a^3}{k_B T} = \frac{2}{3}\frac{a^2}{D_{dif}} \tag{47}$$

in which $D_{dif}$ is diffusion coefficient. Using the parameters of typical molecule (for example, haemoglobin) we obtain $\tau=10^{-7}$s. So, for electromagnetic field of frequency 1MHz or lower, eq (29) is valid. In fact, in alternating field the dielectric constant $\varepsilon$ is dependent of frequency. Define complex permittivity $\varepsilon' = \varepsilon + i\varepsilon^*$ with the real part $\varepsilon$ called dielectric constant and the imaginary part $\varepsilon^*$ dielectric loss. In alternating field we can deduce [25]

$$\varepsilon'(\omega) = \varepsilon_\infty + \frac{\Delta\varepsilon}{1 - i\omega\tau} \quad (\Delta\varepsilon = \varepsilon_s - \varepsilon_\infty)$$

where $\varepsilon_s$ and $\varepsilon_\infty$ are the dielectric constant at zero frequency and at the highest frequency limit respectively. The real part of the equation gives the frequency dependence of dielectric constant

$$\varepsilon(\omega) = \varepsilon_\infty + \frac{\varepsilon_s - \varepsilon_\infty}{1 + \omega^2\tau^2} \tag{48}$$

As $\omega\tau \ll 1$ one has $\varepsilon(\omega) = \varepsilon_s$. If the dispersions of the molecular properties of the cytoplasm are considered there will occur several relaxation terms and eq (48) should be replaced by

$$\varepsilon(\omega) = \varepsilon_\infty + \sum_j \frac{\Delta\varepsilon_j}{1+(\omega\tau)^{2(1-\alpha_j)}} \tag{49}$$

where $\alpha_j$ is the frequency distribution of a certain molecular dispersion. When the frequency is lower than 10MHz the molecular relaxations can be neglected[32]..

4. In the above discussion we proved that the membrane loss $Q_m$ is a small quantity as compared with cytoplasm loss $Q_p$ and can be neglected in electric field case. However, due to the extremely small relative volume a significantly higher entropy production per unit volume within the membrane may have other biological effects[33]. Especially the high frequency field interaction with cell membrane is a class of important events which may be useful in anticancer treatment. The locally applied strong electric field which destabilizes cell membranes in the presence of a drug has been used in electrochemotherapy [34] and has proved effective in skin cancer [35]. Recently, nanosecond pulsed electric field therapy was reported as a novel anti-tumor treatment [36]. In fact, even the low-frequency low-intensity electromagnetic field can cause the marked change of trans-membrane potentials and several mV change of trans-membrane potentials can lead to calcium-ion influxing. The change of intracellular calcium concentration induces the change of cellular behavior, including the apoptosis of cells. Studies have showed that the release of calcium from chondria induce cells apoptosis [37]. The change of ion concentration may influence the signal transduction in a cell and provides an opportunity for anticancer therapy.

5. Electromagnetic fields have been used in cancer treatment for many years. However, the entropy production caused by static magnetic field has not been included in the above discussion Consider the static gradient magnetic field coupling with diffusion flow and insert the magnetic force

$$\mathbf{F} = \frac{1}{\mu_0}\frac{\chi}{\rho n_0}\mathbf{B}\cdot\nabla\mathbf{B} = \frac{1}{\mu_0}\frac{\chi}{\rho n_0}(B_x\frac{\partial}{\partial x}+B_y\frac{\partial}{\partial y}+B_z\frac{\partial}{\partial z})\mathbf{B} \tag{50}$$

into Eq (7). Set the diamagnetic components of a cell denoted by γA and the paramagnetic components denoted by γP. The work completed by magnetic field is

$$\sum_\gamma \mathbf{j}_\gamma\cdot\mathbf{F}_\gamma = \sum_{\gamma A}\mathbf{j}_{\gamma A}\cdot\mathbf{F}_{\gamma A} + \sum_{\gamma P}\mathbf{j}_{\gamma P}\cdot\mathbf{F}_{\gamma P} = (\sum_{\gamma A}\mathbf{j}_{\gamma A})\cdot(\overline{\mathbf{F}_{\gamma A}}-\overline{\mathbf{F}_{\gamma P}})$$

In the last equation we have replaced the forces exerted on diamagnetic or paramagnetic components by their average. For healthy and cancerous cells we have

$$\sum_\gamma (\mathbf{j}_\gamma\cdot\mathbf{F}_\gamma)^{(H)} = (\sum_{\gamma A}\mathbf{j}_{\gamma A}^{(H)})\cdot(\overline{\mathbf{F}_{\gamma A}}-\overline{\mathbf{F}_{\gamma P}}) \cong \lambda n_A^{(H)}$$

$$\sum_\gamma (\mathbf{j}_\gamma\cdot\mathbf{F}_\gamma)^{(C)} = (\sum_{\gamma A}\mathbf{j}_{\gamma A}^{(C)})\cdot(\overline{\mathbf{F}_{\gamma A}}-\overline{\mathbf{F}_{\gamma P}}) \cong \lambda n_A^{(C)} \tag{51}$$

Here $n_A^{(C)}$ and $n_A^{(H)}$ denote the diamagnetic molecule number in unit volume for cancerous cell and healthy cell respectively, λ is a proportionality constant (>0). The relative magnitude of $n_A^{(C)}$ and $n_A^{(H)}$ can be estimated through

$$(n_A^{(C)} - n_A^{(H)})(\chi_A - \chi_P) = N(\chi^{(C)} - \chi^{(H)}) < 0$$

($N = n_A^{(C)} + n_A^{(H)}$). Here $\chi_A$ ($\chi_P$) is the magnetic susceptibility of diamagnetic (paramagnetic) components and $\chi^{(C)}$ and $\chi^{(H)}$ the magnetic susceptibility of cancerous cell and healthy cell respectively, $\chi^{(C)} < \chi^{(H)}$, for example, $\chi^{(C)} = -7.76$, $\chi^{(H)} = -7.16$ for human throat[38]. So, $n_A^{(C)} > n_A^{(H)}$, and from (51) we have

$$\sum_\gamma (\mathbf{j}_\gamma \cdot \mathbf{F}_\gamma)^{(H)} < \sum_\gamma (\mathbf{j}_\gamma \cdot \mathbf{F}_\gamma)^{(C)} \qquad (52)$$

This means the cancerous cell has higher entropy production than the normal under static magnetic field. The introduction of static magnetic field is no helpful in the reversal of entropy flow. But the gradient magnetic field forces are exerted to the diamagnetic (most of organisms) and paramagnetic (mainly the Fe-containing haemoglobin) components of a cancerous cell in opposite directions. A malignant tumour in the fast growing period attracts many new mini blood vessels and get enough alimentation and oxygen. The metabolism and alimentation supply of cancers could be affected when they are placed in the gradient magnetic field [38].

**4. Ultrasound irradiation causes reversal of entropy flow**

Ultrasound absorption in biological tissue leads to additional entropy production in the cells of the tissue. The entropy production rate of a cell $\int \sigma_s^{(5)} d\tau$ under the irradiation of ultrasound is related to ultrasound absorption coefficient $\alpha$ (defined by the decaying of sound intensity $I_s = I_{s0} \exp(-\alpha x)$ where $x$ is the penetrating depth of ultrasound in tissue) through following equation

$$\int \sigma_s^{(5)} d\tau \text{ (ultrasound)} = \alpha D A I_{s0} / T \qquad (53)$$

(where $I_{s0}$ means ultrasonic intensity, $T$ –temperature and $D$ and $A$ – cell diameter and average cross section respectively). Note that the ultrasound absorption does not include the sound attenuation due to scattering in medium, so the absorption rate is related to entropy production directly. We shall calculate absorption coefficient $\alpha$ first.

There are three main mechanisms on ultrasound absorption in biological tissue, namely, the absorption due to the shearing motions of medium molecules and viscous forces, the heat losses due to conduction and the absorption due to chemical relaxation process. The former two are often referred to as classical absorption and the third as non-classical absorption. The chemical relaxation occurs when the equilibrium constant of chemical reaction is affected by pressure changes and/or temperature changes. Corresponding to three kinds of absorption mechanism the absorption coefficient $\alpha$ can be expressed as the sum of three terms

$$\alpha = \alpha_\eta + \alpha_\kappa + \alpha_{chm} \qquad (54)$$

It was proved that [39] the ultrasonic absorption coefficient due to viscous forces

$$\alpha_\eta = \omega^2 \tau_\eta / c_s = 4\eta \omega^2 / (3\rho c_s^3) \qquad (55)$$

and the ultrasonic absorption coefficient due to thermal conduction

$$\alpha_\kappa = \omega^2 \tau_\kappa / c_s = \omega^2 \kappa_T (\gamma_h - 1)/(\rho c_s^3 C_p) \qquad (56)$$

where $\rho$, $\eta$, $C_P$, $\kappa_T$ and $\gamma_h$ are the density, the coefficient of viscosity, the constant-pressure specific heat, the coefficient of heat conductivity and the ratio of isobaric and isothermal heat capacities of the medium, respectively, and $c_s$ is the sound speed and $\omega$ its angular frequency. The sound absorption is a relaxation process. Here $\tau_\eta$ and $\tau_\kappa$ in Eq (55) and (56) mean the mechanical and thermal relaxation time respectively. The total absorption $\alpha$ is related to the equivalent relaxation time $\Gamma$ as follows [40]

$$\alpha = \frac{2\omega}{c_s} \sin \frac{\omega \Gamma \ln \gamma_h}{2\sqrt{1+\omega^2 \Gamma^2}} \tag{57}$$

For low frequency ($\omega\Gamma$ smaller than or near 1), it leads to

$$\alpha = \omega^2 \Gamma \ln \gamma_h / c_s \tag{58}$$

Now we shall make numerical estimates on the sound absorption in biological tissue. Taking ultrasound frequency 1MHz and

$$\eta = 0.015,\ \rho = 1,\ c_s = 1.56 \times 10^5,\ \gamma_h - 1 = 7 \times 10^{-4},\ C_p = 4.18 \times 10^7,$$

$$\kappa_T = 5.88 \times 10^4 \quad \text{(all in CGS unit)} \tag{59}$$

and inserting into Eq (55) and (56) one has

$$\alpha_\eta = 2 \times 10^{-4}\ \text{cm}^{-1} \tag{60}$$

$$\alpha_\kappa = 1.0 \times 10^{-8}\ \text{cm}^{-1} \tag{61}$$

Next, by use of

$$\ln \gamma_h \approx \gamma_h - 1 = 7 \times 10^{-4},\ \Gamma = 0.91 \times 10^{-6}\ \text{s (taken from kidney tissue,[40])}, \tag{62}$$

and Eq. (58) we estimate the values of total absorption $\alpha$

$$\alpha = 1.61 \times 10^{-1}\ \text{cm}^{-1} \tag{63}$$

So

$$\alpha_\eta \ll \alpha \quad \text{and} \quad \alpha_\kappa \ll \alpha \tag{64}$$

We therefore conclude that the most important contribution to the low-frequency ultrasound absorption ($\alpha$) comes from the chemical relaxation ($\alpha_{chm}$).

The experimental data on equivalent relaxation time $\Gamma$ changes largely from tissue to tissue[40]. For example, 0.91 $\mu s$ for kidney, 0.64 $\mu s$ for heart, 0.44 $\mu s$ for testis, 0.15 $\mu s$ for brain, 0.13 $\mu s$ for liver, 0.10 $\mu s$ for tendon and 0.06 $\mu s$ for human blood. Using experimental data on relaxation time $\Gamma$ and eqs (57)(58) we obtain absorption coefficient $\alpha$ for different tissues. Consider liver cell as an example. Inserting

$$\alpha = 0.23 \times 10^{-1}\ \text{cm}^{-1}, D = 30\ \mu m,\ A = \pi D^2/8,$$

$$I_{s0}=0.1 \text{watt/cm}^2 \text{（supposed arbitrarily）}, \quad T=310 \tag{65}$$

into eq (53) we obtain the numerical estimation of the entropy production rate for a cell due to ultrasound absorption

$$\int \sigma_s^{(5)} d\tau \text{ (ultrasound)} = 8\times 10^{-7} \text{ erg/degree/sec} \tag{66}$$

This is in the range of cell entropy production without external field.

Consider the difference of ultrasound entropy production rate between cancer cell and normal cell. Since the chemical relaxation gives the main contribution to the sound absorption and the equilibrium constant of chemical reaction is dependent on pH of cellular environment, the difference of two kinds of cells in ultrasonic absorption is induced mainly by their difference in pH value. In fact, for beef liver supernatant under 1MHz ultrasound irradiation the experiments showed that $\alpha$ increases monotonously with pH in the range from 6 to 12, $\alpha$ =0.1dB/cm for pH=6.4 to $\alpha$ =0.2 dB/cm for pH=12 [41]. The result is understandable since the sound absorption relaxation occurs when the equilibrium constant is affected by pressure changes ($\frac{\partial \ln K_{eq}}{\partial p} = -\frac{\Delta V}{RT}$) and/or temperature changes ($\frac{\partial \ln K_{eq}}{\partial T} = \frac{\Delta H}{RT}$). While $K_{eq}$ is dependent on cell acidity, growing with the pH value, the relaxation time $\Gamma$ should increase with pH, too. As we know, the pH of normal cell (from 7.35 to 7.45) is higher than that of cancer (6.85 to 6.95). This leads to the healthy cell has stronger ultrasonic absorption. Assuming that the relative difference in $\alpha$ between these two cells is 10% shown by experimental data, we estimate the difference of ultrasound entropy production rate between cancer cell and normal cell is $8\times 10^{-7}$ erg/degree/sec for ultrasound power 1 watt/cm$^2$ (see eq (66)).

In the previous section we estimated that for a cell in case of no field applied, the entropy production rate is $4\times 10^{-7}$ erg/degree/sec or lower for cancer and $2\times 10^{-7}$ erg/degree/sec or lower for normal. So, the ultrasound irradiation of intensity 1 watt/cm$^2$ or weaker is capable of making the entropy production rate of normal cell higher than cancer and reversing the direction of entropy flow.

The application of ultrasound technique in anticancer therapy has a long history. In the forties of last century the clinic application of ultrasound had shown curative possibilities. In 1944, when Horvath firstly used ultrasound technique to human tumor treatment, after three days of the operation of ultrasound at 800 KHz, he found that the cancer cells had crushed up. In later years, Horvath conveyed the ultrasound wave to water to irradiate the tumor and gained favorable effect. Several illustrations about the cure of skin cancer by ultrasound had also been reported [42]. In the nineties of last century the high intensity focused ultrasound (HIFU) raised as an effective technique of ablating carcinomas. Although HIFU is an effective and relatively safe treatment several complications have been observed following treatment of certain tumour types using this modality[43]. Simultaneously, the ultrasound at lower frequencies has also been used to enhance drug penetration into the tissue by cavitation.

After ultrasound absorbed in biological media the energy of sound wave is transformed to disordered thermal energy (or in thermodynamic term, entropy) in cells. When the thermal energy accumulated in a cell cannot be effectively dissipated into its environment the temperature of the cell raises. Due to the histographic defect of tumour the blood supply of cancerous cell is always insufficient. The quantity of blood circulation in tumour is only 2% to 15% of surrounding healthy tissue. So, under the strong ultrasound irradiation the thermal dissipation of cancerous cell is more difficult than normal and this makes the temperature of tumour higher than surrounding healthy tissue. This result is helpful in HIFU therapy. However, for the low intensity and low frequency ultrasound irradiation the ultrasonic absorption in biological tissue and thermal energy accumulation in cell is very weak and the induced change of temperature is not important. In this case one should consider the effect of entropy production itself and compare the entropy production rates of two kinds of cells under ultrasound irradiation. We firstly indicated that the higher entropy production rate of normal cell under ultrasound irradiation possibly change the original direction of entropy flow and avoid the propagation of harmful information of cancer into normal tissue. We suggest that this entropy mechanism possibly provides a basis for a novel approach to physiotherapy of anticancer. In fact, the success of low frequency and low intensity ultrasound therapy in several cases of cancer were reported in past years[42]. Perhaps, they could be explained by the entropy mechanism proposed here.

## 5. Conclusions

1. Through the use of general theory of non-equilibrium thermodynamics the entropy production due to various dissipation mechanisms, namely, entropy productions due to temperature differences, chemical potential gradient, viscous stress，chemical affinity are quantitatively calculated for healthy and cancerous cells respectively. It was demonstrated that for different sizes of cells in human body, the rate of entropy production of cancerous cells is always higher than that of (about two times of) healthy cells in case of no external energy input.

2. The information quantity is the projection of thermodynamic entropy from the microscopic phase-space to its subspace spanned by macroscopic variables that describe the chemical, morphological, structural and physiological state of the cell. The entropy flow is the carrier of the information flow. The entropy flow from a normal to a cancerous cell carries the information of the healthy cell, while the entropy flow in the opposite direction carries the harmful information of the cancerous cell. Therefore, it is expected that the harmful effect brought about by the entropy flow from cancerous to healthy tissue can be blocked by the reversal of direction of entropy current through some specially devised mechanism which changes the relative entropy production rates between cancerous and healthy cells.

3. The first mechanism proposed in the article is the modification of cell's pH. Due to the equilibrium constant $K_{eq}$ dependent on pH the standard free energy of the main energy reaction and therefore the entropy production of a cell is dependent on pH. We demonstrated that the decreasing acidity in cancerous cells and/or the increasing acidity in normal cells will change the relative magnitude of entropy production rate of two kinds of cells and cause the reversal of entropy flow.

4. The second mechanism is the electromagnetic field energy input. We demonstrated that several volt/cm low-frequency alternating electric field or $10^4 Gs$ alternating magnetic field with moderate frequency can effectively induce the additional entropy production the order of which is comparable with the entropy production for a cell without applied field. Define the product of conductivity and inverse permittivity square of a cell as its entropy production threshold (EPT). When the EPT (EPT1 or EPT2) of cancerous cell is lower than that of healthy cell then the entropy flow (information flow) direction can be reversed as the organism irradiated under appropriate alternating electromagnetic field. The reversal of entropy flow will avoid the propagation of harmful information of cancer into normal tissue.

5. The third mechanism is introduction of ultrasound irradiation. Through the calculation of ultrasound-induced entropy production and the comparison of theoretical results with experimental data on ultrasound absorption in biological tissues, we demonstrated that, on exposure to low–frequency（<1MHz） low-intensity（<1W/cm$^2$） ultrasound irradiation, the ultrasound absorption will increase the entropy production in normal tissue more efficiently than in tumors due to the more acidic nature of the latter and the direction of entropy flow (information flow) between these two kinds of cells should be reversed. This entropy mechanism provides a basis for a novel non-damage approach to anticancer therapy.

Acknowledgement. The author would like to thank Prof Molnar for numerous discussions on the thermodynamic approach to anticancer. He is also grateful to Dr CJ Ding for discussion on electromagnetic-field-induced entropy production. The work was supported by the National Science Foundation of China, project no. 90403010.